\documentstyle[12pt]{article}
\begin{document}
\centerline{\large\bf $K^+\to\pi^+\nu\bar{\nu}$ with four generations}
\vskip 0.5truecm
\centerline{\large\bf and its effects on unitarity triangle}
\baselineskip=8truemm
\vskip 2.5truecm
\centerline{Toshihiko Hattori$,^{a),}$\footnote{e-mail: 
hattori@ias.tokushima-u.ac.jp} \ Tsutom Hasuike$,^{b),}$\footnote{e-mail: 
hasuike@anan-nct.ac.jp} \ and \ Seiichi Wakaizumi$ ^{c),}$\footnote{e-mail: 
wakaizum@medsci.tokushima-u.ac.jp}}
\vskip 0.6truecm
\centerline{\it $ ^{a)}$Institute of Theoretical Physics, University of Tokushima,
Tokushima 770-8502, Japan}
\centerline{\it $ ^{b)}$Department of Physics, Anan College of Technology,
Anan 774, Japan}
\centerline{\it $ ^{c)}$School of Medical Sciences, University of Tokushima,
Tokushima 770-8509, Japan}
\vskip 2.5truecm
\centerline{\bf Abstract}
\vskip 0.7truecm

We analyze in the four-generation model the first measurement of the 
branching ratio of rare kaon decay $K^+\to\pi^+\nu\bar{\nu}$, using
the constraints from $\Delta m_K, \varepsilon_K, B_d-\bar{B_d}$ mixing,
$\Gamma(b\to s\gamma), B_s-\bar{B_s}$ mixing, $D^0-\bar{D^0}$ mixing, 
$B(K_L\to\pi^0\nu\bar{\nu})$ and $B(K_L\to\mu\bar{\mu})$, and 
study its effects on the unitarity triangle. With the results of searching for the 
maximum mixing of the fourth generation, we predict that $D^0-\bar{D^0}$ 
mixing $\Delta m_D$ and the branching ratio of "direct" CP-violating decay 
process $K_L\to\pi^0\nu\bar{\nu}$ could attain the values 1-2 orders of 
magnitude larger than the predictions of the Standard Model.
\newpage

Recently, the branching ratio of the flavor-changing neutral current(FCNC) 
process, $K^+\to\pi^+\nu\bar{\nu}$, has been measured for the first time 
by the USA-Japan-Canada Collaboration at the Brookhaven National Laboratory, 
and it has turned out to be $B=(4.2^{+9.7}_{-3.5})\times 
10^{-10}$\cite{Adler}. The central value seems to be 4-6 times larger than 
the predictions of the Standard Model $B=(0.6-1.5)\times 10^{-10}$
\cite{Buchalla}.

This process had already been studied by Gaillard and Lee in 1974 and they 
obtained a branching ratio of $\sim 10^{-10}$ by using the "short-distance" 
$W-W$ box and $Z^0$-penguin diagrams in the "4-quark" model\cite{Gaillard}.
After that in 1981, Inami and Lim obtained the rigorous expressions for these 
and other related diagrams relevant to the FCNC processes and studied the effects 
of superheavy quarks and leptons in $K_L\to \mu\bar{\mu}, K^+\to\pi^+\nu
\bar{\nu}$ and $K^0-\bar{K^0}$ mixing\cite{Inami}, before the top-quark 
is discovered.

In this work, we analyze the new branching ratio of $K^+\to\pi^+\nu
\bar{\nu}$ in the four-generation model\cite{Bigi}, since the above-mentioned 
factor 4-6 of the new measurement 
seems to imply the existence of a fourth generation with roughly the same 
mixing as for the third generation. We will search for the maximum mixing  
for the "hypothetical" fourth generation by imposing the constraints from 
$\Delta m_K, \varepsilon_K, B_d-\bar{B_d}$ mixing, 
$\Gamma(b\to s\gamma), B_s-\bar{B_s}$ mixing, $D^0-\bar{D^0}$ mixing, 
$B(K_L\to\pi^0\nu\bar{\nu})$ and $B(K_L\to\mu\bar{\mu})$, and study 
its effects on CP violation in neutral $B$ meson decays and the unitarity triangle.

For the unitary $4\times 4$ quark mixing matrix, we will use the 
Hou-Soni-Steger parametrization\cite{Hou}, which has a simple form in the 
third column; $(V_{ub}, V_{cb}, V_{tb}) = (s_zc_ue^{-{\rm i}\phi_1}, 
s_yc_zc_u, c_yc_zc_u)$, in the fourth row; $(V_{t'd}, V_{t's}, V_{t'b}, 
V_{t'b'}) = (-c_uc_vs_we^{{\rm i}\phi_3},$ 

\noindent
$-c_us_ve^{{\rm i}\phi_2}, -s_u, c_uc_vc_w)$ and $V_{us} = s_xc_zc_v - 
s_zs_us_ve^{{\rm i}(\phi_2-\phi_1)}$, where the three mixing angles 
$s_x(\equiv\sin\theta_x),  s_y$ and $s_z$ give the elements $|V_{us}|, 
|V_{cb}|$ and $|V_{ub}|$, respectively as in the Standard Model, the 
phase $\phi_1$ corresponds to the Kobayashi-Maskawa(KM) CP-violating 
phase $\delta^{KM}$\cite{Kobayashi}, and $s_u, s_v$ and $s_w$ are the 
new mixing angles and $\phi_2$ and $\phi_3$ are the new phases, $t'$ and 
$b'$ being the fourth generation up- and down-quark, respectively.

As an input, we use the following values\cite{Buchalla} 
\begin{equation}
s_x=0.22, \qquad s_y=0.040\pm 0.003, \qquad s_z/s_y=0.08\pm 0.02,  
\label{shiki1}
\end{equation}
in the same way as in the Standard Model, since the magnitude of the three 
elements $V_{us}, V_{cb}$ and $V_{ub}$ are experimentally determined 
from the semileptonic decays of hyperons and $B$ mesons, and the existence 
of a fourth generation would not affect the determination.We search for the 
maximum mixing of the fourth generation by testing the three cases of 
$(s_w,s_v,s_u)=(\lambda^4,\lambda^3,\lambda^2), (\lambda^3,\lambda^2,
\lambda)$ and $(\lambda^2,\lambda^2,\lambda)$\cite{Hayashi}, 
where $\lambda\equiv
0.22 \simeq\sin\theta_C$ is the expansion parameter used in the Wolfenstein 
parametrization of the $3\times 3$ KM matrix. The constraints we impose on 
the model are the following; $K_L-K_S$ mass difference $\Delta m_K=(3.522
\pm 0.016)\times 10^{-12}$ MeV\cite{Particle}, CP-violating parameter 
in the neutral kaon system $\varepsilon_K=(2.28\pm 0.02)\times 10^{-3}$
\cite{Particle}, $\Delta m_{B_d}=(3.12\pm 0.20)\times 10^{-10}$ MeV
\cite{Particle} for $B_d-\bar{B_d}$ mixing, $B(b\to s\gamma)=(2.32\pm 
0.67)\times 10^{-4}$\cite{Alam} for the inclusive radiative $b$ decay, 
$B(K^+\to\pi^+\nu\bar{\nu})=(4.2^{+9.7}_{-3.5})\times 10^{-10}
$\cite{Adler}, $x_s>10.5$\cite{Adam} for $B_s-\bar{B_s}$ mixing strength, 
$\Delta m_D<1.4\times 10^{-10}$ MeV\cite{Aitala} for $D^0-\bar{D^0}$ 
mixing, $B(K_L\to\pi^0\nu\bar{\nu})<5.8\times 10^{-5}$\cite{Weaver} and 
$B(K_L\to\mu\bar{\mu})_{\rm SD}<4.4\times 10^{-9}$, where the 
short-distance(SD) contribution to $B(K_L\to\mu\bar{\mu})$ is taken to be 
the value two times larger than the one by B\'{e}langer and Geng\cite{Belanger} 
as a loose constraint.

Each of the above-mentioned nine constraints is studied in the following.
\vskip 0.3truecm

\noindent
(i)$K_L-K_S$ mass difference, $\Delta m_K$

\noindent
The short-distance part of $\Delta m_K$ comes from the well-known $W-W$ 
box diagram with $c, t$ and $t'$ as internal quarks and the contribution is expressed, 
for example, for the box with two $c$-quarks as follows,
\begin{equation}
\Delta m_K(c,c)=\frac{G_F^2M_W^2}{12\pi^2}f_K^2B_Km_K{\rm Re}
[(V_{us}V_{cd}^*)^2]\eta^K_{cc}S(x_c),  \label{shiki2}
\end{equation}
where $S(x)$ is the Inami-Lim box function\cite{Inami}, $x_c\equiv m_c^2/
M_W^2$, $m_c$ being the charm-quark mass, $\eta^K_{cc}$ is the QCD 
correction factor including the next-to-leading order effects, and $f_K$ and $B_K$ 
are the decay constant and bag parameter of the kaon, respectively. By taking 
for these parameters the values of $m_c=1.3$ GeV, $\eta^K_{cc}=1.38$
\cite{Buchalla}, $f_K=0.16$ GeV and $B_K=0.75\pm 0.15$\cite{Buchalla}, 
we obtain from the inputs of eq.(\ref{shiki1}) the $(c,c)$ contribution $\Delta 
m_K(c,c)=(2.6-3.9)\times 10^{-12}$ MeV, which is already consistent by itself 
with the measured value. Numerically, the $(c,t)$ and $(t,t)$ contributions are 
very small as compared with the $(c,c)$ contribution, so we take a constraint for 
the fourth-generation contributions to be 
\begin{equation}
\left| \frac{\Delta m_K(c,t')+\Delta m_K(t,t')+\Delta m_K(t',t')}
{\Delta m_K(c,c)}\right| <1      \label{shiki3}
\end{equation}
as a loose constraint, since there are a large amount of long-distance 
contributions.
\vskip 0.3truecm

\noindent
(ii)CP-violating parameter in neutral kaon system, $\varepsilon_K$

\noindent
The quantity $\varepsilon_K$ is expressed by the imaginary part of hadronic matrix 
element of the effective Hamiltonian with $\Delta S=2$ between $K^0$ and 
$\bar{K^0}$, to which the short-distance contribution comes from the $W-W$ 
box diagram as in $\Delta m_K$. The box contribution with $c$ and $t$ quarks 
gives an expression of
\begin{equation}
\varepsilon_K(c,t)=\frac{1}{\sqrt{2}\Delta m_K}
\frac{G_F^2M_W^2}{6\pi^2}f_K^2B_Km_K{\rm Im}[V_{cs}V_{cd}^*
V_{ts}V_{td}^*]\eta^K_{ct}S(x_c,x_t).     \label{shiki4}
\end{equation}
If we take the QCD correction factor including the next-to-leading order 
as $\eta_{ct}^K$= $0.47$ \cite{Buchalla}, the dominant term in the 
$(c,t)$-box contribution leads to $\varepsilon_K(c,t)\simeq 2.83
\times 10^{-3}B_K\sin\phi_1$ 
for $m_t=180$ GeV, where $\phi_1$ is the CP-violating phase. Since this 
magnitude of $\varepsilon_K(c,t)$ is close to the measured value, we take the 
constraint from $\varepsilon_K$ that the sum of the contributions from $c, t$ 
and $t'$ quarks should be within the $1\sigma$ error of the measured value,
\begin{equation}
\sum_{i,j=c,t,t',i\le j}\varepsilon_K(i,j)=(2.28\pm 0.02)\times 10^{-3}.
\label{shiki5}
\end{equation}
The theoretical uncertainty in the bag parameter $B_K=0.75\pm 0.15$ is taken 
into consideration.
\vskip 0.3truecm

\noindent
(iii)$B_d-\bar{B_d}$ mixing, $\Delta m_{B_d}$

\noindent
The mass difference between the two mass-eigenstates of $B_d-\bar{B_d}$ 
system is given by the $W-W$ box diagram, and the $(t,t)$-box contribution 
is expressed by
\begin{equation}
\Delta m_{B_d}(t,t)=\frac{G_F^2M_W^2}{12\pi^2}f_B^2B_Bm_{B_d}
\left| V_{tb}V_{td}^*\right| ^2\eta^B_{tt}S(x_t),   \label{shiki6}
\end{equation}
where $f_B$ and $B_B$ are the decay constant and the bag parameter for 
$B_d$ meson, respectively, and $\eta^B_{tt}$ is the QCD correction factor 
including the next-to-leading order effects. By taking for these parameters the 
values of ${\sqrt B_B}f_B=(0.20\pm 0.04)$ GeV\cite{Buchalla} and 
$\eta^B_{tt}=0.55$\cite{Buchalla} and by using the inputs of eq.(\ref{shiki1}), 
we obtain the $(t,t)$ contribution; $\Delta m_{B_d}(t,t)=(1.75-3.95)\times 
10^{-10}$ MeV, of which range includes the measured value. Since $(c,c)$ and 
$(c,t)$ contributions are numerically very small as compared with the $(t,t)$ 
contribution, we take the constraint from $\Delta m_{B_d}$ 
that the sum of the contributions from $t$ and $t'$ should be within the 
$1\sigma$ error of the measured value, $\Delta m_{B_d}=(3.12\pm 
0.20)\times 10^{-10}$ MeV\cite{Particle}. 
\vskip 0.3truecm

\noindent
(iv)$B(b\to s\gamma)$

\noindent
The dominant contribution to the inclusive radiative $b$ decay, $b\to s\gamma$, 
comes from the electromagnetic penguin diagram with $t$- and $t'$-quark exchange 
in the four-generation model. The partial decay width is given by\cite{Hewett}
\begin{equation}
\Gamma (b\to s\gamma)=\frac{\alpha G_F^2m_b^5}{128\pi^4}\left|V_{tb}
V_{ts}^*c_7(m_b)+V_{t'b}V_{t's}^*c'_7(m_b)\right|^2,   \label{shiki7}
\end{equation}
where $\alpha$ is the fine-structure constant and $c_7(m_b)$ and $c'_7(m_b)$ are 
the Wilson coefficients for the electromagnetic dipole operator, calculated via 
leading-logarithmic evolution equation with the electromagnetic penguin functions 
at the electroweak scale\cite{Inami} down to the renormalization scale $\mu =m_b
(=4.5$ GeV)\cite{Grinstein} for the $t$- and $t'$-exchange diagrams, respectively.
We take the constraint from $B(b\to s\gamma)$ that the sum of $t$ and $t'$ 
contributions to the decay width of eq.(\ref{shiki7}) should be within the 
1$\sigma$ error of $\Gamma (b\to s\gamma)=(9.54\pm 2.76)\times 10^{-17}$ 
GeV, calculated from the branching ratio and the lifetime of $B_d$ meson, 
$\tau_{B_d}=1.60$ ps\cite{Particle}.
\vskip 0.3truecm

\noindent
(v)$B(K^+\to \pi^+\nu\bar{\nu})$

\noindent
The short-distance contributions to the rare decay $K^+\to\pi^+\nu\bar{\nu}$ are
from the $W-W$ box diagram and $Z^0$-penguin diagram. The branching ratio is 
given by\cite{Buchalla}
\begin{equation}
B(K^+\to\pi^+\nu\bar{\nu})=\kappa_+\left| \frac{V_{cd}V_{cs}^*}{\lambda}
P_0+\frac{V_{td}V_{ts}^*}{\lambda^5}\eta_tX_0(x_t)+\frac{V_{t'd}
V_{t's}^*}{\lambda^5}\eta_{t'}X_0(x_{t'})\right|^2,      \label{shiki8}
\end{equation}
where $\kappa_+=4.57\times 10^{-11}$, $P_0$ is the sum
of charm contributions to the two diagrams including the next-to-leading order QCD 
corrections\cite{Buras}, $X_0(x_t)$ and $X_0(x_{t'})$ the sum of the $W-W$ box
and $Z^0$-penguin functions for $t$- and $t'$-quark exchange\cite{Inami}, 
respectively, $\eta_t(=0.985)$ is the next-to-leading order QCD correction to the 
$t$-exchange 
calculated by Buchalla and Buras\cite{Buras93} and we will take $\eta_{t'}=1.0$ 
for $t'$-exchange. The constraint is that the branching ratio of eq.(\ref{shiki8}) 
should be consistent with the measured value of branching ratio $B=(4.2^{+9.7}
_{-3.5})\times 10^{-10}$\cite{Adler}, since the long-distance contribution is 
estimated to be very small $(B\sim 10^{-13})$\cite{Rein}. We do not assume 
the mixing in the leptonic sector.
\vskip 0.3truecm

\noindent
(vi)$B_s-\bar{B_s}$ mixing, $x_s$

\noindent
The dominant contribution to $B_s-\bar{B_s}$ mixing is the $W-W$ box diagram
with $t$- and $t'$-exchange as in $B_d-\bar{B_d}$ mixing. We take the constraint 
that the sum of $(t, t), (t, t')$ and $(t', t')$ contributions to the  mixing strength 
should be larger than the present experimental lower bound $x_s>10.5$ 
\cite{Adam}, where $x_s\equiv \Delta m_{B_s}/\Gamma_{B_s}$, 
$\Delta m_{B_s}$ being the mass difference of the two mass eigenstates of 
$B_s-\bar{B_s}$ system.
\vskip 0.2truecm

\noindent
(vii)$D^0-\bar{D^0}$ mixing, $\Delta m_D$

\noindent
The dominant contribution to $D^0-\bar{D^0}$ mixing in the four-generation 
model is the $W-W$ box diagram with fourth-generation down-quark $b'$ 
exchange\cite{Babu}. We take the constraint that this contribution to the mass 
difference between the two mass-eigenstates of $D^0-\bar{D^0}$ system 
should be smaller than the present experimental upper bound\cite{Aitala},
$\Delta m_D(b',b')<1.4\times 10^{-10} {\rm MeV}$, 
since the Standard Model box contribution of two $s$-quarks exchange
\cite{Datta}and the long-distance contributions\cite{Wolfenstein} are estimated 
to be three to four orders of magnitude smaller than the upper bound.
\vskip 0.3truecm

\noindent
(viii)$B(K_L\to \pi^0\nu\bar{\nu})$

\noindent
The process $K_L\to \pi^0\nu\bar{\nu}$ is a "direct" CP-violating decay
\cite{Littenberg} and the rate is expressed by the imaginary part of sum of the 
same $W-W$ box and $Z^0$-penguin diagram amplitudes as in $K^+\to \pi^+
\nu\bar{\nu}$\cite{Buchalla}. We take the constraint that the sum of $t$ and 
$t'$ contributions to the branching ratio should be smaller than the experimental 
upper bound\cite{Weaver} $B(K_L\to\pi^0\nu\bar{\nu})<5.8\times 
10^{-5}$. 
\vskip 0.3truecm

\noindent
(ix)$B(K_L\to\mu\bar{\mu})_{\rm SD}$

\noindent
The process $K_L\to\mu\bar{\mu}$ is a CP-conserving decay. The 
short-distance(SD) contribution is given by the $W-W$ box and 
$Z^0$-penguin diagrams and the branching ratio for this part is expressed as
\cite{Buchalla}
\begin{equation}
B(K_L\to\mu\bar{\mu})_{\rm SD}=\kappa_{\mu} \left[ \frac{{\rm Re}
\left( V_{cd}V_{cs}^*\right) }{\lambda}P'_0+\frac{{\rm Re}\left( V_{td} 
V_{ts}^*\right) }{\lambda^5}Y_0(x_t)+\frac{{\rm Re}\left( V_{t'd}
V_{t's}^*\right) }{\lambda^5}Y_0(x_{t'})\right]^2,      \label{shiki9}
\end{equation}
where $\kappa_{\mu}=1.68\times 10^{-9}$, $P'_0$ the sum 
of charm contributions to the two diagrams including the next-to-leading order 
QCD corrections\cite{Buras} and $Y_0(x_t)$ and $Y_0(x_{t'})$ are the sum 
of the $W-W$ box and $Z^0$-penguin functions for $t$- and $t'$-exchange, 
respectively\cite{Inami}. We take the constraint that the branching ratio of 
eq.(\ref{shiki9}) should be smaller than the upper bound of the short-distance 
contribution as stated before, $B(K_L\to\mu\bar{\mu})_{\rm SD}<4.4
\times 10^{-9}$. 

In order to find the maximum mixing for the fourth generation consistent with 
the above nine constraints, we study the following three cases of $(\left| V_{t'd}
\right| ,\left| V_{t's}\right| ,\left| V_{t'b}\right| )\simeq (s_w,s_v,s_u)=
(\lambda^4,\lambda^3,\lambda^2), (\lambda^3,\lambda^2,\lambda)$ and
 $(\lambda^2,\lambda^2,\lambda)$, where $\lambda=0.22$ is the Cabibbo 
angle. We tentatively take the mass of the fourth generation quarks $(t',b')$ 
as $m_{t'}=400$ GeV and $m_{b'}=350$ GeV so as to satisfy the constraints 
obtained from the analyses with the oblique parameters $S, T$ and $U$
\cite{Particle}\cite{Kawakami}.
\begin{table}
\caption{Combinations of relevant quark mixing matrix elements for 
$\Delta m_{B_d}, b\to s\gamma, K^+\to\pi^+\nu\bar{\nu}$ and 
$(K_L\to\mu\bar{\mu})_{\rm SD}$ for the third generation and the three 
cases of fourth generation mixing.}
\begin{center}
\begin{tabular}{ccccc}  \hline\hline
Mixing & $\Delta m_{B_d}$ & $b\to s\gamma$ & $K^+\to\pi^+\nu
\bar{\nu}$ & $(K_L\to\mu\bar{\mu})_{\rm SD}$ \\  \hline
$(V_{td},V_{ts},V_{tb})$ & $V_{td}V_{tb}$ & $V_{ts}V_{tb}$ & 
$V_{td}V_{ts}$ & $V_{td}V_{ts}$ \\
$(\lambda^3,\lambda^2,1)$ & $\lambda^3$ & $\lambda^2$ & $\lambda^5$ 
& $\lambda^5$ \\  \hline
$(V_{t'd},V_{t's},V_{t'b})$ & $V_{t'd}V_{t'b}$ & $V_{t's}V_{t'b}$ & 
$V_{t'd}V_{t's}$ & $V_{t'd}V_{t's}$ \\
$(\lambda^4,\lambda^3,\lambda^2)$ & $\lambda^6$ & $\lambda^5$ & 
$\lambda^7$ & $\lambda^7$ \\
$(\lambda^3,\lambda^2,\lambda)$ & $\lambda^4$ & $\lambda^3$ & 
$\lambda^5$ & $\lambda^5$ \\
$(\lambda^2,\lambda^2,\lambda)$ & $\lambda^3$ & $\lambda^3$ & 
$\lambda^4$ & $\lambda^4$ \\  \hline\hline
\end{tabular}
\end{center}
\label{tab1}
\end{table}
\vskip 0.1truecm

Strong constraints come from $\Delta m_K, \varepsilon_K, B_d-\bar{B_d}$ 
mixing, $b\to s\gamma, K^+\to\pi^+\nu\bar{\nu}$ and $(K_L\to\mu
\bar{\mu})_{\rm SD}$. In the Standard Model, the largest contribution comes 
from the top-quarks for $B_d-\bar{B_d}$ mixing, $b\to s\gamma, K^+\to
\pi^+\nu\bar{\nu}$ and $(K_L\to\mu\bar{\mu})_{\rm SD}$, and the 
combination of the relevant quark mixing matrix elements is $V_{td}V_{tb}
\sim\lambda^3$ for $B_d-\bar{B_d}$ mixing, $V_{ts}V_{tb}\sim
\lambda^2$ for $b\to s\gamma$, and $V_{td}V_{ts}\sim\lambda^5$ for 
$K^+\to\pi^+\nu\bar{\nu}$ and $(K_L\to\mu\bar{\mu})_{\rm SD}$.
The combinations of the corresponding matrix elements for $t'$-quark are 
shown in Table 1 for each of the above three cases. By comparing these 
combinations between the Standard Model and the four-generation model, 
the numerical analyses give the following results;the case of $(\lambda^4,
\lambda^3,\lambda^2)$ gives almost the same predictions to the 
above-mentioned nine processes as in the Standard Model and the contributions
of the fourth generation are very small. So, this case is not interesting. 
For the case of $(\lambda^3,\lambda^2,\lambda)$, almost all the processes 
satisfy the constraints with only one exception of $B(K_L\to\mu
\bar{\mu})_{\rm SD}$, for which this mixing gives a value almost seven 
times larger than the upper bound. The last case of
$(\lambda^2,\lambda^2,\lambda)$ predicts too large values for $B(K^+\to
\pi^+\nu\bar{\nu})$ and $B(K_L\to\mu\bar{\mu})_{\rm SD}$.
These results imply that the mixing $(\lambda^3,\lambda^2,\lambda)$ is 
a little large for the fourth generation and it turns out that a mixing with $s_w$ 
and $s_v$ reduced by $20\%$, that is, $(s_w,s_v,s_u)= (0.8\lambda^3,
0.8\lambda^2,\lambda)$ satisfies all of the nine constraints as a maximum 
mixing.
\begin{table}
\caption{Comparison of $B(K^+\to\pi^+\nu\bar{\nu}), x_s(B_s-\bar{B_s} 
{\rm mixing}), \Delta m_D$ and $B(K_L\to\pi^0\nu\bar{\nu})$ among  
the experimental values, Standard Model(SM) predictions and four-generation 
model predictions with maximum mixing.}
\begin{center}
\begin{tabular}{lcccc}  \hline\hline
& $B(K^+\to\pi^+\nu\bar{\nu})$ & $x_s$ & $\Delta m_D({\rm MeV})$ 
& $B(K_L\to\pi^0\nu\bar{\nu})$  \\  \hline
Experiment & $\left( 4.2^{+9.7}_{-3.5}\right) \times 10^{-10}$ & $>10.5$ 
& $<1.4\times 10^{-10}$ & $<5.8\times 10^{-5}$  \\
SM & $(0.6-1.5)\times 10^{-10}$ & $19-27$ & $\sim 10^{-14}$ 
& $(1.1-5.0)\times 10^{-11}$  \\
4-generation & $(0.7-4.4)\times 10^{-10}$ & $19-29$ & $(0.7-2.1)
\times 10^{-12}$ & $(0.05-10)\times 10^{-10}$  \\  \hline
\end{tabular}
\end{center}
\label{tab2}
\end{table}
\vskip 0.1truecm

We can obtain the following predictions from this maximum mixing; the 
branching ratio of $K^+\to\pi^+\nu\bar{\nu}$ takes a range from the 
Standard Model(SM) values to the central value of the new measurement as $B=
(0.7-4.4)\times 10^{-10}$, the strength for $B_s-\bar{B_s}$ mixing 
is $19\le x_s\le 29$, $\Delta m_D$ of $D^0-\bar{D^0}$ mixing could have 
a value $(0.7-2.1)\times 10^{-12}$ MeV, about two orders of magnitude 
larger than the SM prediction ($\sim 10^{-14}$ MeV\cite{Datta}), 
and the branching ratio of $K_L\to\pi^0\nu\bar{\nu}$ takes a range 
of $(0.05-10)\times 10^{-10}$, 
from  the SM values to the ones two orders of magnitude larger than the SM 
prediction ($(1.1-5.0)\times 10^{-11}$\cite{Buchalla}). These results are 
summarized in Table 2.
The branching ratios of $K^+\to\pi^+\nu\bar{\nu}$ and $K_L\to\pi^0\nu
\bar{\nu}$ are correlated with each other as shown in Fig.1 for the maximum 
mixing, the area of the correlation resulting from the freedom of the three 
phases $\phi_1, \phi_2$ and 
$\phi_3$. In the region of $1.2\times 10^{-10}\le B(K^+\to\pi^+\nu
\bar{\nu})\le 4.5\times 10^{-10}$, the correlation is around the line $B(K_L
\to\pi^0\nu\bar{\nu})=4.5B(K^+\to\pi^+\nu\bar{\nu})-1.3\times 
10^{-10}$, which is caused by the positive collaboration of third and fourth 
generations. On the other hand, in the region of $0.7\times 10^{-10}\le 
B(K^+\to\pi^+\nu\bar{\nu})\le 1.2\times 10^{-10}$, an interference 
of second- and third-generation contributions with fourth-generation ones 
brings this range of SM values of $B(K^+\to\pi^+\nu\bar{\nu})$ and 
a broad range of $B(K_L\to\pi^0\nu\bar{\nu})=(0.05-4)\times 10^{-10}$.

The maximum mixing gives an interesting effect on CP-asymmetry of the decay 
rates of the "gold-plate" mode of $B_d$ meson, $B_d\to J/\psi K_S$. The 
asymmetry is given by 
\begin{equation}
C_f=\frac{\Gamma (B_d\to J/\psi K_S)-\Gamma (\bar{B_d}\to J/\psi K_S)}
{\Gamma (B_d\to J/\psi K_S)+\Gamma (\bar{B_d}\to J/\psi K_S)},     
\label{shiki10}
\end{equation}
and it is expressed as\cite{Carter}
\begin{equation}
C_f=-\frac{x_d}{1+x_d^2}{\rm Im}\Lambda, \qquad  \Lambda\equiv
{\sqrt \frac{M_{12}^*}{M_{12}}}\frac{A(\bar{B_d}\to J/\psi K_S)}
{A(B_d\to J/\psi K_S)},    \label{shiki11}
\end{equation}
where $x_d$ is the mixing strength for $B_d-\bar{B_d}$ mixing, $M_{12}$ 
the off-diagonal element of the mass matrix in $B_d-\bar{B_d}$ system 
and $A$ is the decay amplitude. In the Standard Model\cite{Dunietz}, 
the quantity $C_f$
takes a positive sign as $0.18\le C_f\le 0.37$, which results from the phase 
range $0<\phi_1<\pi$, constrained from the positive sign of $\varepsilon_K$. 
However, in the four-generation model\cite{Hasuike}, $C_f$ can take also 
a negative sign as $-0.38\le C_f\le 0.40$, since the phase $\phi_1$ can take 
the whole range of $0<\phi_1<2\pi$ due to the two more new phases 
$\phi_2$ and $\phi_3$ and the maximum mixing of the fourth generation. 
Although in the four-generation model the penguin diagrams could affect 
the decay amplitude, they would cause at most several percent change of 
the value of $C_f$, even if they happen to have a magnitude as large 
as $50\%$ of the tree amplitude.

Second, the unitarity triangle in the Standard Model transforms into unitarity 
quadrangle in the four-generation model\cite{Nir}. For the maximum mixing 
obtained here, some of the typical quadrangles are shown in Fig.2. The fourth 
side of the quadrangle, $V_{t'd}V_{t'b}^*$, is of order $\lambda^4$, while 
the other three sides are of order $\lambda^3$. The first example of Fig.2(a) is
for positive value of $C_f$. The second one of Fig.2(b) is for negative value 
of $C_f$ and the quadrangle is reversed with respect to the base line of 
$V_{cd}V_{cb}^*$, since $\phi_1>\pi$.

Summarizing, we find a maximum mixing of the fourth generation $(V_{t'd},
V_{t's},V_{t'b})\simeq (0.8\lambda^3, 0.8\lambda^2, \lambda)$, which is 
consistent with the nine constraints from $\Delta m_K, \varepsilon_K, 
B_d-\bar{B_d}$ mixing, $b\to s\gamma, K^+\to\pi^+\nu\bar{\nu}, 
B_s-\bar{B_s}$ mixing, $D^0-\bar{D^0}$ mixing, $K_L\to\pi^0
\nu\bar{\nu}$ and $K_L\to\mu\bar{\mu}$. The mass difference 
$\Delta m_D$ from $D^0-\bar{D^0}$ mixing and the branching ratio of
$K_L\to\pi^0\nu\bar{\nu}$ could reach the values two orders of 
magnitude larger than the Standard Model predictions, and CP asymmetry 
of the decay rates of $B_d\to J/\psi K_S$ could take a value of opposite 
sign to the SM one. 

We are grateful to Takeshi Komatsubara, Minoru Tanaka, Takeshi Kurimoto, 
Xing Zhi-Zhong, Masako Bando, C.S. Lim, and Morimitsu Tanimoto for 
helpful discussions.

\newpage
\centerline{\bf Figure captions}

\vskip 1.0truecm
\noindent
{\bf Fig.1.} The correlation of $B(K^+\to\pi^+\nu\bar{\nu})$ and $B(K_L
\to\pi^0\nu\bar{\nu})$ for the maximum mixing $(s_w,s_v,s_u)=
(0.8\lambda^3,0.8\lambda^2,\lambda)$ in the four-generation model. The 
hatched area is the allowes region for the branching ratios. The rectangle 
surrounded by  the dashed lines is the prediction of the Standard Model. 

\vskip 0.5truecm
\noindent
{\bf Fig.2.} Typical examples of the unitarity quadrangle. (a) $\phi_1=\frac{1}
{3}\pi, \phi_2=\frac{11}{6}\pi, \phi_3=\frac{19}{12}\pi; C_f(B_d\to 
J/\psi K_S)=0.39, 
B(K^+\to\pi^+\nu\bar{\nu})=2.1\times 10^{-10}$, (b) $\phi_1=\frac{19}
{12}\pi, \phi_2=\pi, \phi_3=\frac{3}{4}\pi; C_f(B_d\to J/\psi K_S)=-0.35, 
B(K^+\to\pi^+\nu\bar{\nu})=2.7\times 10^{-10}$.
\end{document}